\documentclass{iopart}   
\usepackage{graphicx}   
\usepackage{dcolumn}    

\begin{document}

\title{The falling chain of Hopkins, Tait, Steele and Cayley }
\author{Chun Wa Wong, Seo Ho Youn and Kosuke Yasui}
\address{Department of Physics and Astronomy, 
University of California, Los Angeles, CA 90095-1547, USA}
\ead{cwong@physics.ucla.edu}   

\begin{abstract}
A uniform, flexible and frictionless chain falling link by link from a heap by the edge 
of a table falls with an acceleration $g/3$ if the motion is nonconservative, but $g/2$ 
if the motion is conservative, $g$ being the acceleration due to gravity. Unable to 
construct such a falling chain, we use instead higher-dimensional versions of it. 
A home camcorder is used to measure the fall of a three-dimensional version called 
an $xyz$-slider. After frictional effects are corrected for, its vertical falling 
acceleration is found to be $a_x/g = 0.328 \pm 0.004$. This result agrees with the 
theoretical value of $a_x/g = 1/3$ for an ideal energy-conserving $xyz$-slider. 
\end{abstract}

\maketitle

\section{Introduction}
\label{sect:Introduction}

A uniform, flexible and frictionless chain falling link by link from a heap by the edge 
of a table is a popular example or problem that appears in many textbooks of classical 
mechanics \cite{Wong06}. The solution given in these books is almost always the 
energy-nonconserving one where the chain falls with an acceleration of $a=g/3$, $g$ 
being the acceleration due to gravity. This solution was given as early as 1857 by 
Cayley \cite{Cayley57,Irschik04,Wong06}. Recently, Villarino \cite{Villarino06} has
called our attention to a footnote in Routh's treatise on dynamics \cite{Routh98} 
that refers to problems of this type involving systems with variable masses being 
solved in the private lecture room of the famed Cambridge tutor Hopkins as long ago 
as 1850. The footnote also mentions a solution to our falling-chain problem published 
in 1856 in a treatise on dynamics by Tait and Steele \cite{Tait56}. For this reason, 
we shall call the chain falling from a heap at the table edge the 
Hopkins-Tait-Steele-Cayley (HTSC) falling chain.

Mikhailov \cite{Mikhailov84} has given a history of dynamics problems involving variable 
masses, especially those that like the HTSC falling chain were made up for or studied 
by undergraduates in Cambridge in the second half of the 19th century in connection 
with the Cambridge Mathematical Tripos Examination \cite{Warwick03}. It thus seems useful 
to extend the review given in Wong and Yasui \cite{Wong06} on the HTSC chain in order to 
clarify the historical circumstances under which these dynamics problems were studied in 
Cambridge. This extended review will be given in section~\ref{sect:TSHchain}.

Wong and Yasui \cite{Wong06} have taken issue with the energy-nonconserving solution 
of $g/3$ for the acceleration of the HTSC chain. They point out that the full Lagrange 
equation of motion contains an extra term not given by the solution of Cayley and 
of Tait and Steele. This missing term gives the chain tension with which the falling 
chain is pulling on the falling link. When this term is included, the solution becomes 
energy-conserving and the falling acceleration is increased to $g/2$. The difference 
in physics between the energy-conserving and nonconserving solutions will also be 
described in section~\ref{sect:TSHchain}.

\begin{figure}
\includegraphics{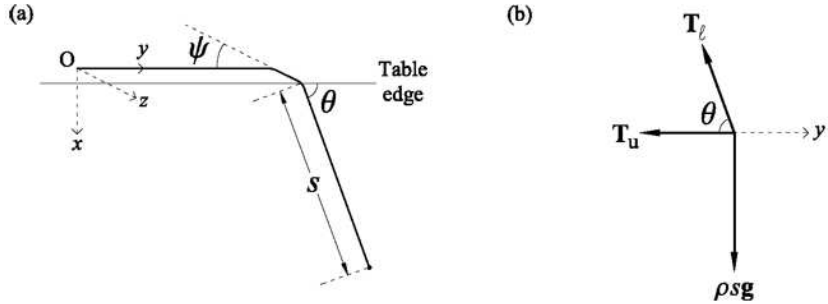}
\caption{\label{fig:xySlider}
(a) The falling chain called the $xy$-slider that has one end fixed at the origin O
of the coordinate system. The chain slides along the table edge after release. The 
falling chain is called an $xyz$-slider when the stationary segment on the table parallel 
to the edge is set back from the edge by a distance $z = d$. (b) The forces acting on 
the chain bend of the $xy$-slider at a sharp table edge. } 
\end{figure}
It is thus of interest to determine experimentally which of these two solutions for the 
HTSC falling chain is correct. We immediately run into the difficulty that a heap or coil 
of chain on a table cannot be built in a reproducible way, for every heap seems unique. 
After much thought, we discover a two-dimensional realization of the falling chain called 
the $xy$-slider that is easily reproducible. This is a 
chain stretched out on the table parallel to the table edge and very close to it, as 
shown schematically in figure~\ref{fig:xySlider}(a). One end of the chain is held fixed 
on the table with the fingers of one hand while the other end overhangs the table 
slightly. When the overhanging end is allowed to fall, the chain is found to slide along 
the slightly round table edge to the fixed end, as shown in the figure. The reason for 
holding down one end is to ensure that the resting `heap' on the table (now represented 
by a straight chain segment parallel to the table edge) remains stationary on the table 
while the rest of the chain falls. It also makes the chain less susceptible to 
perturbations caused by the release of the falling end. The chain motion then becomes more 
easily reproducible. 

The motion of a uniform, flexible and frictionless $xy$-slider is described in 
section~\ref{sect:xySliderMotion}, using Lagrangian mechanics. Such an ideal chain 
experiences no dissipation so that its energy is conserved during the chain fall. 
For the special case where the fallen chain segment is a straight line at a constant 
angle $\theta$ to the horizontal table edge, the theoretical falling acceleration $a$ 
is found to be constant in time, with  
\begin{eqnarray}
\theta &=& \cos^{-1}(1/3) = 70.5^\circ, \nonumber \\
a &\equiv& \ddot{s} = g/\sqrt{8} = 0.354 g, \nonumber \\
a_x &=& a\sin\theta = g/3,
\label{thetaAs}
\end{eqnarray}
where $s$ is the length of the fallen chain segment.

We now need to determine experimentally how an actual $xy$-slider falls. In the rest of 
the paper, we shall show that the required measurements can be made using only simple 
equipment found at home, so that the interested reader can readily repeat our measurement. 
We shall begin, in section~\ref{sect:FreeFall}, by showing that an 
inexpensive home camcorder \cite{camcorder} recording at 30 frames a second has sufficient 
resolution to time the fall of chains. We verify its capability by measuring the known 
value of $g$ for the free fall of a metal object. The measurement on the $xy$-slider is 
finally described in section~\ref{sect:xySliderResult}. We explain why a three-dimensional 
version of the slider called an $xyz$-slider is used. The results roughly corrected for 
friction in the air, on the table, and at the table edge are 
\begin{eqnarray}
\theta &=& 70^\circ \pm 2^\circ, \nonumber \\
a_x/g &=& a \sin\theta/g = 0.328 \pm 0.004, \nonumber \\ 
a/g &=& 0.350 \pm 0.006.
\label{thetaAs}
\end{eqnarray}
The corrected motion of a real $xyz$-slider is thus found to agree with that of the 
ideal energy-conserving $xy$-slider. Our result thus gives indirect experimental 
support that the ideal HTSC chain is also energy conserving, and can be expected to 
have a falling acceleration of $g/2$.

\section{The history and physics of the HTSC falling chain}
\label{sect:TSHchain}

\begin{figure}
\includegraphics{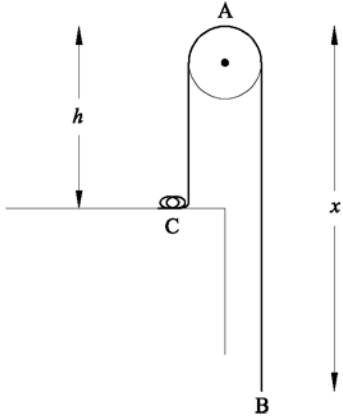}
\caption{\label{fig:TSChain}
The falling chain of Tait and Steele \cite{Tait56}. } 
\end{figure}

According to Mikhailov \cite{Mikhailov84}, rocket propulsion by the continuous emission 
of masses was studied as early as 1810 by Moore. A Lagrange equation of motion for a more 
general mechanical system with variable mass was given by Buquoy in 1812, and re-written 
in a more modern form by Poisson in 1819. We are interested here in the variable-mass 
problem of a flexible and frictionless chain falling link by link from a heap at 
the table edge. This problem is a special case of the following 
problem described and solved by Tait and Steele in a textbook on classical dynamics 
published in 1856 \cite{Tait56,Warwick03,Villarino06}: `One end, B, of a uniform heavy 
chain hangs over a small smooth pulley A, and the other end is coiled up on a table at 
C. If B preponderates, determine the motion.' This falling chain is shown in 
figure~\ref{fig:TSChain}. 

Tait and Steele solve the problem by using Newton's force law written in the form
\begin{eqnarray}
\frac{d}{dt}[\rho (x+h)v] = \rho g (x-h),
\label{NewtonEq}
\end{eqnarray}
where $\rho$ is the uniform linear mass density of the chain, $x$ is the length {\it AB}, 
the chain end as measured from the pulley, $h$ is the length {\it AC}, the height of the 
pulley above the table, and $v = dx/dt$. By multiplying both sides of the equation by 
$(x+h)v$, (\ref{NewtonEq}) can be integrated immediately to give 
\begin{eqnarray}
{\textstyle \frac{1}{2}}(x+h)^2v^2 = {\textstyle \frac{g}{3}}(x-x_0)
(x^2 + x_0x + x_0^2-3h^2),
\label{vsq}
\end{eqnarray}
if initially $x=x_0$ and $v=0$. If $x_0=2h$ also holds, the acceleration becomes 
constant in time, $a = g/3$. Our falling HTSC chain corresponds to the special case 
$h=x_0=0$ of this problem, the case of a bare chain without the pulley.

If the total mechanical energy of the falling chain is conserved, however, the 
result would have been $v^2 = gx$, with $a = g/2$. Sommerfeld in his textbook on 
mechanics \cite{Sommerfeld52} calls the missing mechanical energy $\Delta E = \rho x^2g/6$ 
Carnot's energy loss to emphasize the picture that the missing energy is lost when a 
falling link from the table falls by sticking onto the falling chain in an inelastic 
collision. Inelastic impacts where the masses stick together after the impact are known 
to lose total mechanical energy, a result first described by Lazare 
Carnot in a work published in 1783 \cite{Sommerfeld52,Ball60,Gillispie71}.

A year after the publication of the treatise of Tait and Steele, Cayley wrote a 
paper \cite{Cayley57} in which the same equation of motion (\ref{NewtonEq}) is obtained 
from the Lagrangian 
\begin{eqnarray}
{\cal L}(x,v) = \frac{\rho}{2} xv^2 +  \frac{\rho g}{2} x^2.
\label{Lagrangian}
\end{eqnarray}
We are unable to follow Cayley's arguments, and shall just describe his final result.
His version of Lagrange's equation contains the same term describing the rate of momentum 
change as the left-hand side of (\ref{NewtonEq}), but the part where the variable mass 
of the falling chain is differentiated is obtained in a different way with the help of 
another function (denoted $K$ in his paper). His prescription is such that in calculating 
the generalized force
\begin{eqnarray}
\frac{\partial {\cal L}(x,v)}{\partial x} = \rho gx + T,
\label{Wt&T}
\end{eqnarray}
the $x$ dependence of the variable mass of the first (or kinetic-energy) term of the 
Lagrangian is not to be differentiated. The consequence is that the Wong-Yasui tension 
$T = \rho v^2/2$ that comes from such a differentiation does not appear. As a result, 
Cayley's equation agrees with (\ref{NewtonEq}) of Tait and Steele. Cayley nevertheless 
considers his method more general. Mikhailov \cite{Mikhailov84} has pointed out,
however, that Cayley's unusual prescription of selective differentiation has not been 
adopted by any subsequent worker in the field.

It is interesting that Cayley did not mention the Tait-Steele solution in his paper. 
Cayley was very knowledgeable, of course, about theoretical developments in dynamics. 
In fact, earlier in 1857 he wrote a rather detailed and very well-received 42-page 
report on theoretical dynamics that summarized the progress of the subject since 
Lagrange's {\it M\'ecanique Analytique} of 1788. In this report, Cayley discussed the 
work of Lagrange, Poisson, Hamilton, Jacobi, Liouville, Bertrand, Denkin and others, 
up to early 1857 \cite{Cayley57BA,Crilly06}, all by name. We believe that the probable 
reason Cayley did not refer to Tait and Steele is that both the problem and its solution 
were so well known in Cambridge circles that a reference to Tait and Steele seemed 
unnecessary, and perhaps even unjustified. 

In this connection, it is worth pointing out that according to Routh \cite{Routh98}, 
problems of impact or infinitesimal impulse like that attributed to our falling chain 
had been studied as early as 1850 in the private lecture room of William Hopkins, a 
popular private tutor (`coach' in Cambridge undergraduate slang \cite{Warwick03a}) to 
Cambridge undergraduates preparing for the Mathematical Tripos examination. Cayley, Tait 
and Routh were all coached by Hopkins, who helped them win the highest score in the 
Mathematical Tripos of 1842, 1852 and 1854, respectively \cite{Warwick03b}. (The highest 
scorer in the Mathematical Tripos each year was known for life as the Senior Wrangler, 
Wranglers being holders of first-class Cambridge B.A. degrees in mathematics.) Since
Routh was a student in Hopkins's lecture room before the Mathematical Tripos Exam of 
1854, the date 1850 given in his footnote on Hopkins's work on variable-mass problems 
can be considered reliable.

On the other hand, Cayley took the Tripos exam eight years earlier in 1842. So it 
appears likely that he did not, like Routh and Tait, learn about this class of 
problems with variable masses directly from Hopkins in the latter's lecture room. We 
should add that after a four-year fellowship at the Trinity College, Cayley left 
Cambridge in 1846 to study law in London \cite{Forsyth95,Crilly98}. However, Cayley in 
his London days was in touch with Cambridge. He served as a Cambridge college 
examiner in the years 1848-51 \cite{Forsyth95}. In 1852, he was elected a fellow of 
the Royal Society with the help of many Cambridge supporters, including William 
Thompson (Lord Kelvin), Stokes and Hopkins \cite{Crilly98}. He also served as the 
Senior Examiner of the Mathematical Tripos that year. Tait, the Senior Wrangler that 
year, became one of his friends \cite{Forsyth95}. Cayley's 1857 paper, received by 
the journal on 18 June, 1857, begins with the sentence `There are a class 
of dynamical problems which, so far as I am aware, have not been considered 
in a general manner.' His paper was therefore written about a year after 
the publication of the Tait and Steele book on dynamics \cite{Tait56}, where Tait's 
Preface was dated January, 1856, and written from Queen's College in 
Belfast where he was the professor of mathematics since 1854. We give all these 
historical details in order to suggest that Cayley did not acknowledge Tait and
Steele perhaps because the problem and the solution were not invented by them. 
We are not sure if this is true, because we have not been able to find any 
earlier reference on this falling-chain problem. 

Wong and Yasui \cite{Wong06} have given a list of other textbooks on mechanics where 
the energy-nonconserving solution for the HTSC falling chain can be found. Since the 
problem is so closely connected to the Cambridge Mathematical Tripos Examination, 
it is interesting to add to this list of books the 1878 second 
edition of Wolstenholme's popular problem collection \cite{Wolstenholme78} for the 
Mathematical Tripos exams, Jeans's textbook on theoretical mechanics \cite{Jeans07} 
and Lamb's book on dynamics \cite{Lamb14}. Wolstenholme (Third Wrangler of 1850), Lamb 
(Second Wrangler of 1872 and a student of Routh) and Jeans (Third Wrangler of 1898) all 
took the Mathematical Tripos Examination \cite{Warwick03e}. 

\section{Motion of the $xy$-slider}
\label{sect:xySliderMotion}

We have noticed that the $xy$-slider shown in figure~\ref{fig:xySlider}(a) falls in a
very special way. Soon after release, the falling chain assumes an equilibrium shape 
in which a rather straight falling chain segment slides along the table edge at a 
constant angle $\theta$. Depending on the chain and on the manner of release, there are 
in general some transverse vibrations and distortions from the linear shape. We shall 
limit ourselves to the simple case where these complications are either absent or small, 
and the fallen chain segment is approximately a straight line of length $s$ at an angle 
$\theta$ to a {\it sharp} table edge, as illustrated in the figure. For a chain of 
length $L$, the coordinates of the falling chain end as seen from the fixed end O are 
\begin{eqnarray}
(x,y,z) = (s \sin\theta, L - s + s\cos\theta, 0).
\label{ChEnd}
\end{eqnarray}
Using the two-dimensional form $(\rho s/2)(\dot{x}^2 + \dot{y}^2)$ of the kinetic energy 
of the fallen chain segment, we find the two-dimensional Lagrangian 
\begin{eqnarray}
{\cal L}_2(s,v=\dot{s}) = \rho s v^2 (1-\cos\theta) + \frac{\rho g}{2}s^2\sin\theta,
\label{Lagrangian}
\end{eqnarray}
where $\rho$ is the uniform linear mass density of the chain, and $v = \dot{s}$. 
There is no contribution from the stationary chain segment on the table as its kinetic 
energy vanishes, and its potential energy can be taken to be zero. Our model is thus 
greatly simplified from the general case of two-dimensional motion,  for which 
each term of the Lagrangian is an integral in $ds$. 

The two-dimensional Lagrangian (\ref{Lagrangian}) is, like its
one-dimensional counterpart, not an explicit function of the time $t$. Hence the total 
mechanical energy of the system is conserved. If initially $s=0$ and $v = 0$, then
\begin{eqnarray}
v^2 = \frac{g}{2} \frac{\sin\theta}{1-\cos\theta}s.
\label{vsq2}
\end{eqnarray} 
Such a velocity comes from the constant acceleration 
\begin{eqnarray}
a \equiv \ddot{s} = \frac{g}{4} \frac{\sin\theta}{1-\cos\theta} 
\label{a}
\end{eqnarray} 
along the longitudinal direction of the falling chain segment. These solutions for $v^2$ 
and $a$ satisfy the Lagrange equation of motion 
\begin{eqnarray}
\left(2s\ddot{s} + \dot{s}^2\right)(1-\cos\theta) - sg\sin\theta = 0.
\label{LagrangeEq}
\end{eqnarray} 

To understand the preferred direction $\theta$ of the $xy$-slider, we consider 
first a plumb line whose support moves horizontally with the same acceleration 
$g_y' = d^2(L-s)/dt^2 = - a$ as the bend of the chain at the table edge. In the 
laboratory frame, this plumb line does not point vertically downward, but at an 
angle $\theta$ to the table edge such that
\begin{eqnarray}
\cot\theta = \frac{a}{g} = \frac{\sin\theta}{4(1-\cos\theta)}, \quad {\rm or} \nonumber \\
(3\cos\theta - 1)(\cos\theta - 1) = 0.
\label{cosines}
\end{eqnarray}
The solution $\cos\theta = 1$ or $\theta = 0$ is unphysical because it requires 
$a = \infty$. The physical solution is $\theta =  \cos^{-1}(1/3) = 70.5^\circ$. 

It can be expected intuitively that the chain actually falls along the modified plumb 
direction ${\bf g}'$ found here. To show that this is indeed the case, we concentrate on 
the variable-mass chain segment of length $s$ that has already fallen below the table 
edge. If the chain bend at the table edge is sharp, the fallen chain segment is acted on 
simultaneously by its own downward weight, an upper chain tension ${\bf T}_u$ along the 
negative $y$-axis, and a lower tension ${\bf T}_\ell$ (if any) pointing up at an angle 
$\theta$ above the horizon, as shown in figure~\ref{fig:xySlider}(b). Its Lagrange 
equations in the $x$- and $y$-directions are then
\begin{eqnarray}
\rho sg - T_\ell\sin\theta &=& \frac{d}{dt}(\rho s\dot{x}), 
\label{LaGxEq} \\
-T_u -T_\ell\cos\theta &=&  \frac{d}{dt}(\rho s\dot{y}). 
\label{LaGyEq}
\end{eqnarray}
They can be simplified to the Newton equations 
\begin{eqnarray}
\rho sg - T_N\sin\theta &=& \rho s\ddot{x} = \rho s a\sin\theta, 
\label{xEq} \\
\Delta -T_N\cos\theta &=&  \rho s\ddot{y} = - \rho s a(1-\cos\theta), 
\label{yEq}
\end{eqnarray}
where $\Delta = -T_u + \rho\dot{s}^2$ and the lower {\it Newtonian} tension
\begin{eqnarray}
T_N = T_\ell + \rho\dot{s}^2
\label{TN}
\end{eqnarray}
is an effective force that differs in general from the {\it Lagrangian} tension $T_\ell$.
These Newton equations can be solved for $a$ and $T_N$:
\begin{eqnarray}
a &=& g\cot\theta - \frac{\Delta}{\rho s}, \nonumber \\
T_N &=& \rho sg \frac{1-\cos\theta}{\sin\theta} + \Delta. 
\label{aTN}
\end{eqnarray}

The magnitude $T_u$ of the upper chain tension at the table edge satisfies the Newton 
equation $T_u = \rho L |\ddot{Y}|$, where $Y$ is the $y$-coordinate of the center of mass: 
\begin{eqnarray}
Y = \frac{1}{2L} [L^2 - s^2(1-\cos\theta)].
\label{YCM}
\end{eqnarray} 
Hence
\begin{eqnarray}
T_u &=& 3\rho as (1-\cos\theta), \nonumber \\
\Delta &=& \rho as(3\cos\theta - 1).
\label{upperTension}
\end{eqnarray}     
where use has been made of (\ref{vsq2}). These expressions when inserted into 
(\ref{aTN}) give
\begin{eqnarray}
a &=& \frac{g}{3\sin\theta}, \quad {\rm or} \quad a_x = \frac{g}{3}, \nonumber \\
T_N &=& 2\rho as, \quad {\rm or} \quad T_\ell = 0.
\label{aTN2}
\end{eqnarray}     
We have thus shown that the Lagrangian chain tension vanishes below the chain bend. 
The equation of motion (\ref{LaGxEq}) then becomes identical to that for the incorrect 
energy-nonconserving solution of the one-dimensional HTSC chain. This is why the 
energy-conserving $xy$-slider has the same downward acceleration $g/3$ as the 
incorrect solution of the one-dimensional HTSC chain. We recall that the correct 
result of $g/2$ for the energy-conserving HTSC chain is obtained only with the help of 
a nonzero downward-pointing Lagrangian tension that has been called the Wong-Yasui 
tension in section \ref{sect:TSHchain}. 

The constant acceleration $a$ of the $xy$-slider must also satisfy (\ref{a}) because 
of energy conservation: 
\begin{eqnarray}
a = \frac{g}{3\sin\theta} = \frac{g\sin\theta}{4(1-\cos\theta)}.
\label{aMatching}
\end{eqnarray}
This requirement can be used to show that the angle $\theta$ satisfies the same condition 
(\ref{cosines}) as the plumb line. The chain thus falls along the plumb direction. 
When this happens, the Newtonian chain tension below the sharp bend has the same magnitude
as the tension ${\bf T}_u$ above the bend, namely $T_N = T_u$, while $\Delta$ vanishes.

\section{Testing the experimental setup for free fall}
\label{sect:FreeFall}

We now have to measure experimentally the acceleration of a real $xy$-slider. 
The falling time over a distance $x$ under a constant acceleration $a$ is
\begin{eqnarray}
t = \sqrt{\frac{2x}{a}}.
\label{t}
\end{eqnarray}
For $x=0.5$ m, the falling time is $t=0.32$ (0.45, 0.55) s when 
$a = g\;(g/2,\; g/3).$ For a camcorder recording at 30 frames a second, the event 
lasts 9.5 (13.5, 16.5) frames. Such a camcorder thus appears adequate for our timing
needs.

To test the camcorder as a timer, we first measure the falling time of several identical 
$5/16''$ steel nuts (as in nuts and bolts) over a height of 1.1-1.3 m. A small golden 
plastic bead is tied to each nut to improve its visibility in the video record. The nut 
is initially held between two fingers, one from each hand. At $t=0$, the fingers are 
pulled away horizontally from each other. By extrapolating the recorded finger separations 
in two succeeding frames, we can estimate the starting time to about 0.2 frame, or 0.007 s. 
A convenient final frame is then chosen and the falling distance shown in it is measured. 

The distance measurement can be done by transferring the $2.5''$ LCD screen image to a PC 
using the camcorder manufacturer's software. The metal nuts are dropped just in front 
of a lunch counter. The known height of the counter from table top to the top of the footrest 
both showing in the picture allows the measurements to be converted into actual distances 
in meters. 

The falling positions are measured from the video record by two persons independently. 
They both report a reading uncertainty of 0.004 m. (After the measurements are completed, 
we find that the PC images can be captured by another software and enlarged for more 
accurate readings.) All results and errors reported here are the averages from 
these two independent readings. 

The observed acceleration 
calculated from the total elapsed time turns out to be larger than expected. To find out 
the reason, we measure the falling position $x(t)$ frame by frame in the video recording. 
The data are fitted well by the functional form
\begin{eqnarray}
x(t) = c_0 + c_1t + c_2t^2.
\label{xt}
\end{eqnarray}
The fit gives a small initial downward velocity $c_1 > 0$ in all runs, thereby showing 
that the person releasing the steel nut does not do it cleanly. The acceleration 
due to gravity obtained from $g = 2c_2$ represents the value after correcting for 
the initial velocity. The corrected values are found to be more consistent 
between runs by a factor of 2 than the values obtained from the total elapsed time. 

A simple (unweighted) average over $N=14$ runs gives the 
experimental result of $g \pm \delta g = 9.82 \pm 0.03$ ms$^{-2}$ shown in 
table~\ref{tab:ExptResults}. Here $\delta g$ is the standard deviation for the mean of 
$N$ measurements:
\begin{eqnarray}
(\delta g)^2 = \frac{s_g^2}{N} = \frac{1}{N(N-1)}\sum_{i=1}^N(g_i - g)^2,
\label{sampleStdDev}
\end{eqnarray}
where $s_g$ is the standard deviation for a single measurement. This $\delta g$ contains 
only statistical contributions. No systematic errors will be estimated for free fall. 
The camcorder timer is assumed to have the accuracy $6\times 10^{-6}$ of a quartz 
clock \cite{Sony}.
\begin{table}
\caption{\label{tab:ExptResults}
The vertical acceleration measured for free fall and for the $xyz$-slider. The position 
of the fixed end of the $xyz$-slider is also noted. The measurements on the $xyz$-slider 
are described in section~\ref{sect:xySliderResult}. }
\begin{indented}
\item[]\begin{tabular}{lcc}
\br
 Quantity $a$  & Simple mean $a (\delta a)$ & Weighted mean $a (\delta_0 a, \delta_k a)$ \\
\mr
Free fall $g$            &  9.820 (0.034) &  9.818 (0.017, 0.033)  \\                      
$xyz$-slider $a_x$:       &      &         \\
\quad Left               &  3.149 (0.013)   & 3.147 (0.013, 0.011)  \\
\quad Right              &  3.136 (0.016)   & 3.129 (0.014, 0.014)  \\
\quad Average            &  3.144 (0.010)   & 3.139 (0.009)  \\
\br
\end{tabular}
\end{indented}
\end{table}

A more sophisticated estimate of $g$ is possible. The least-square fit to the frame 
by frame data for each free fall gives a fitting uncertainty $\delta g_{0i}$, taken to 
be the standard deviation or `standard error' calculated by the {\it Mathematica} 
linear-regression function Regress \cite{Wolfram}. The weighted mean and its raw 
standard error are then calculated from the formula \cite{Rosenfeld75}
\begin{eqnarray}
g \pm \delta_0 g = \frac{\sum_iw_ig_i}{\sum_i w_i} \pm (\sum_i w_i)^{-1/2},
\label{weightedMean}
\end{eqnarray}
with the weights $w_i = (\delta g_{0i})^{-2}$. The goodness-of-fit parameter for all $N$ 
measurements,
\begin{eqnarray}
\chi^2 = \sum_iw_i(g_i - g)^2,
\label{chisq}
\end{eqnarray}
turns out to be larger than $N-1$, the number of degrees of freedom. This means that 
the fit is poor, or that the estimated errors are too small. Following standard 
practice \cite{Rosenfeld75}, all errors are next increased by a scale factor
\begin{eqnarray}
k = \left[\frac{\chi^2}{N-1}\right]^{1/2}
\label{kScale}
\end{eqnarray}
so that the re-computed value of $\chi^2/(N-1)$ becomes exactly 1 after the scaling. 
In our final report, we use the larger of the two errors $\delta_0 g$ and 
\begin{eqnarray}
\delta_k g = k\delta_0 g
\label{deltak}
\end{eqnarray}
for any $k$.

Table~\ref{tab:ExptResults} shows that the weighted mean agrees very well with the 
simple mean in both value and uncertainty. Their close agreement occurs because the 
values of both $g_i$ and $\delta g_{0i}$ are clustered together, thus indicating 
that the experimental data are consistent from run to run. The measured value of 
$g = 9.82 \pm 0.03$ ms$^{-2}$ agrees with the known sea-level value at Los Angeles 
(33$^\circ 56'\;N$) of $g = 9.796$ ms$^{-2}$ \cite{HCP91}. The use of our camcorder as 
a timer is thus validated.

\section{Result for the $xyz$-slider}
\label{sect:xySliderResult}

A real falling $xy$-slider differs in many ways from the theoretical idealization 
described in section~\ref{sect:xySliderMotion}. First of all, we need to use a rounded 
table edge to make sure that the chain links round it smoothly. The chain must then 
fall over the rounded table edge link by link rather than collectively. Such a 
controlled fall is achieved by setting the chain back from the rounded table edge by 
a parallel distance $d$. The setback $xy$-slider is thus an 
$xyz$-slider, with the $z$-axis on the table and perpendicular to the table edge, as 
shown in figure~1(a). Initially the chain runs along the $y$-axis from the fixed end at
the origin O. It then turns $90^\circ$ to run along the $z$-axis for a distance $d$ 
before it runs over the table edge to overhang the table top a distance $x_0$ along the 
vertical $x$-axis. On release, the moving chain rapidly assumes a shape shown in the 
figure where the setback segment makes an angle $\psi$ with the table edge, and has length 
\begin{eqnarray}
s_d = \frac{d}{\sin\psi}.
\label{sd}
\end{eqnarray}
The falling chain segment in the $xy$-plane has length $s$ and soon makes an angle 
$\theta$ with the table edge. The length of the stationary chain segment on the table 
along the $y$-axis is $L-s-s_d$, assuming a sharp table edge for simplicity. 

We now show that a setback $d$ of the chain from the table edge can be compensated by a 
proper choice of the initial overhang $x_0$. With a sharp table edge, the chain Lagrangian 
for the $xyz$-slider is
\begin{eqnarray}
{\cal L}_3(s,v) \approx \rho(s + s_d) v^2(1-\cos\theta) + 
 \frac{\rho g}{2} s^2\sin\theta,
\label{Lag3}
\end{eqnarray}
where we have ignored the difference between the angles $\psi$ and $\theta$ and have 
treated the segment $s_d$ as a continuation of $s$ in the kinetic energy. Energy 
conservation can be used to find the longitudinal velocity $v = \dot{s}$ of the fallen 
chain segment:
\begin{eqnarray}
v^2 &\approx& 2a \frac{s^2 - s_0^2}{s + s_d},
\label{xyzECons}
\end{eqnarray}
where $a = (g/4)\sin\theta/(1-\cos\theta)$ is the constant acceleration without setback
or overhang. We see that for the choice $s_0 \equiv x_0/\sin\theta = s_d$, the chain 
falls with the same constant acceleration $a$ as the $xy$-slider without setback.  
(\ref{xyzECons}) can then be written more simply as
\begin{eqnarray}
x - x_0 \approx {\textstyle \frac{1}{2}} (a\sin\theta)t^2,
\label{xx0}
\end{eqnarray}
assuming that the overhang angle $\theta$ is constant in time. 

The sharp table edge used in our theoretical analysis greatly simplifies all formulas,
but it is mathematically different from the rounded table edge used in the experiment. 
In the motion of another chain sliding as a whole
off the table edge, we have studied equations of motion for both sharp and rounded edges 
and shown that while they are mathematically distinct, they give states of motion that 
are very close to each other numerically \cite{Wong06a}. Hence the use of
a sharp table edge in our theoretical analysis is acceptable. 

A real $xyz$-slider also experiences frictional drags as it moves. There are at least 
four sources of friction: (1) between the table surface and the setback chain segment of
length $s_d$ on the table, (2) between the table edge and the chain rounding it, 
(3) between neighboring chain links at the bends both on the table and at the table edge 
(bending friction), and (4) air friction resisting the horizontal and vertical motions of 
the falling chain segment. The influence of both setback and frictional forces can in 
principle be compensated by a proper choice of the initial overhang $x_0$ of the chain. 
Frictional force \#~1 is just the usual 
\begin{eqnarray}
F_1 = \mu_k\rho g s_d,
\label{F1}
\end{eqnarray}  
where $\mu_k$ is the coefficient of kinetic friction. The other frictional effects are 
much more complicated. They will be described in \ref{systematic}. 

We have used a variety of light ball and cable chains in our measurements, each chain 
about a meter long. We have used a steel desk, a marble counter 
top, and construction lumber with edges routed to different radii of curvature. Our most 
consistent results are those where the falling chain shows the least amount of transverse 
distortion and vibrations. They are obtained with a light steel cable chain used 
for making necklaces and jewellery. A small golden plastic bead (of diameter $3/16''$ 
$\approx 0.48$ cm) is tied to the falling end to make it more visible in the camcorder 
picture. Chain and bead together is 0.949 m long, 11.744 g in mass, to which the bead 
contributes 0.0569 g. The chain has 316 links, making the average link 0.299 cm long, 
and 0.0370 g in mass. The bead is about 1.5 chain links both in mass and in length. 

The table edge used in the reported result is made from a piece of $2''\times 6''$ 
construction lumber (hereafter referred to as `the table') with the edge routed by a 
$1/4'' \approx 0.64$ cm (radius) router bit. Both the edge and the wood surface on which 
the chain slides are covered by Teflon tapes \cite{TeflonTape} to reduce friction. The use 
of wet lubricants on our light chains seems to increase friction through surface tension 
rather than to reduce it. The use of dry graphite powder reduces friction 
marginally, but we decide that it is not worth the mess. So all measurements are done 
on a dry unlubricated surface. Our experience is that construction lumber made of fir 
is too soft and grainy to make a good table edge. We recommend using pine or oak 
instead. For a semi-quantitative demonstration, however, the choice of a table edge is 
not critical.

The levelness of the tabletop is adjusted by shimming with folded pieces of paper and 
checked by using an inexpensive level embedded in a heavy 4-foot plastic yardstick. The 
frictional coefficient $\mu_k$ for our chain and surface is measured by placing the chain 
on the table in a straight line perpendicular to the edge and partially overhanging it. 
The overhanging length is then varied until the whole chain begins to move. The resulting 
static frictional coefficient is $\mu_s = 0.16$ without tapping, and 0.15 with gentle 
tapping of the table. The kinetic frictional coefficient can be a complicated function of 
the velocity \cite{Persson00,RoyMech}, but is usually considered to be significantly smaller 
than $\mu_s$ \cite{RoyMech}. We shall use a relatively large value of $\mu_k = 0.13$.

In our measurements, we first arrange the offset segment along the $z$-direction and the
overhang segment in the vertical direction. The chain is held at the bend on the table 
with a pen controlled by hand. The chain is released by rapidly moving the pen upwards in 
a vertical direction. The starting time is determined by linear extrapolation of the 
vertical position of the releasing pen in two succeeding video frames. 

The theoretical `edge' of the rounded table edge is defined to be the line parallel 
to the $y$-axis at an angle of $\phi_0 \approx 23^\circ$ from the upward vertical 
(negative $x$-axis). This angle is 
obtained as follows: A point mass initially at rest at $\phi_0$ moving on a frictionless 
circle that is the rounded edge in cross section will fall off the circle and fall down 
by gravity to the horizontal position of $\phi = \pi/2$. The angle $\phi_0$ is so 
defined that the falling time of the mass is the same as the time of free fall from the 
table top through a vertical distance that is the radius of the circle to the same final 
height.  

In doing the measurements, we feel most comfortable working with an offset of 4 chain 
links, or $d \approx 1.2$ cm. The initial overhang is thus $x_0 = d = 4$ links for the 
initial values of $\theta = \psi = 90^\circ$. The equilibrium shape after release has
$\psi \approx 45^\circ$ as estimated by eye, and an overhang angle of 
$\theta = 70^\circ \pm 2^\circ$ as measured from the video recording. The compensatory 
overhang at the equilibrium shape would have been  $x_0 = d\sin\theta/\sin\psi \approx 6$ 
links. Since we cannot change the overhang once the chain is released, we use a middling 
value of 5 links and add another 0.5 link for friction. After the measurements, a more 
detailed study of frictional effects shows that this frictional correction is too small. 
So an additional correction will be applied to the measured result. 

\begin{figure}
\includegraphics{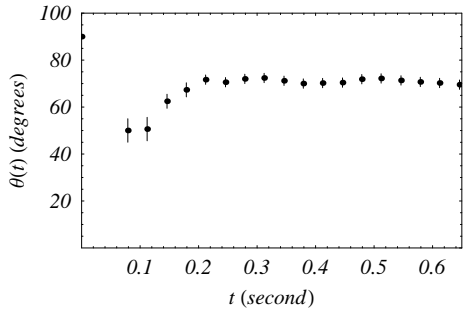}
\caption{\label{fig:Angle}
Time dependence of the overhang angle $\theta(t)$ in one of the runs of the $xyz$-slider.
The initial value of $\theta$ is 90$^\circ$.  }
\end{figure}
On release, the change in chain shape begins at the bend on the table with the angle $\psi$ 
rapidly taking the value of $\approx 45^\circ$. This chain direction extends to below the chain 
bend at the table edge before enough of the chain has fallen down to enforce the equilibrium 
value of $\theta \approx 70^\circ$. The time dependence of $\theta(t)$ for one of the runs 
measured by one reader is shown in figure~\ref{fig:Angle}. We can see an initial transient in 
overhang angle that lasts 0.2 second. The slight modulation of $\theta(t)$ after 0.2 s 
indicates vibration caused by edge imperfections or disturbance on release.  

It is hard to ensure that our chain release is clean. Furthermore, the larger static 
friction acting initially may also affect the result. To reduce these uncertainties, we 
fit the falling distance $x(t)$ frame by frame to the three-term power series (\ref{xt})
used for free fall. This fit corrects specifically for differences 
in initial position and velocity in different runs, while random errors in the fitted 
acceleration itself will be reduced by averaging over several runs. We do 10 runs with the 
left chain end held fixed on the table and 10 runs with the right chain end held fixed, 
to average out any unlevelness of the table. 

A final complication is that the vertical scale of the video image is not exactly the same
from left to right. This change of scale is handled by measuring the vertical falling position 
$x$ in each frame relative to the fiducial distance (between the table top and the top of 
the footrest) at the appropriate $y$-position of the moving chain end seen in that frame. 
The results for $a_x$ from the two independent readers agree well, in fact better than the 
results for free fall. Again, all reported values of $a_x$ are the averages of the results 
from the two readers. The accelerations obtained by fitting $x(t)$ to (\ref{xt}) are 
summarized in table~\ref{tab:ExptResults}. The average of the two sets of runs with 
different fixed end gives $a_x = 3.139 \pm 0.009$ ms$^{-2}$ or $a_x/g = 0.3204 \pm 0.0010$, 
where the uncertainties are statistical only. 

We now turn to systematic corrections and errors. The analysis is rather lengthy, and shall 
be given in \ref{systematic}. The conclusion is that the main source of energy loss is  
air friction on the sliding chain and that the raw acceleration should be increased by 
$2.5\% \pm 1.2 \%$ to the value of $a_x = 3.22 \pm 0.04$ ms$^{-2}$ or $a_x/g = 0.328 
\pm 0.004$, and $a/g = 0.350 \pm 0.006$. We have not included a correction for vibrations 
and other transverse motions of the chain that is likely to bring the result closer 
to the theoretical value of $a_x = g/3$ for the ideal frictionless chain.

Our error analysis shows that some of the refinements used are not necessary in a 
semi-quantitative demonstration since we have obtained quite good results with the many
tables used. Further improvement of the quantitative result obtained will require 
a better correction for air friction, a cleaner method of chain release to reduce chain 
distortion and vibrations, a correction for the energy lost to these transverse 
motions and a better table edge. The effect of the initial transient in chain shape should 
be understood. It would be great to automate the reading of the falling position $x(t)$ 
for it is a rather tedious job. However, our error analysis shows that we have enough 
position resolution. 

\appendix

\section{Systematic errors and corrections}
\label{systematic}

The length of the moving chain segment on the table is $s_d = d = 4$ links initially, but 
$s_d = d/\sin\psi \approx 6$ links on reaching equilibrium shape. A middling value of 5 
links is used for the purpose of data analysis. We estimate that this choice has an 
uncertainty of 0.5 link. The resulting uncertainty on the experimental acceleration 
$a_{\rm exp}$ can be found by writing (\ref{xyzECons}) in the form
\begin{eqnarray}
a_{\rm raw} = a_{\rm exp} \frac{s+s_0}{s+s_d},
\label{aRaw}
\end{eqnarray}
where $a_{\rm raw}$ is the raw acceleration calculated under the assumption of exact 
compensation between overhang and setback. This equation shows that if the overhang $s_0$ 
is too small by an amount $\Delta s_0 = s_d - s_0$, then the measured $a_{\rm raw}$ is 
too small. It should be increased by the fractional amount
\begin{eqnarray}
\frac{\Delta a}{a_{\rm raw}} = \frac{\Delta s_0}{s+s_0}.
\label{Da}
\end{eqnarray}
Taking $s$ to be half of the average falling distance, or $(0.37/\sin\theta)\,{\rm m} = 0.39$ m, 
we find an uncertainty of $\Delta a /a_{\rm raw} = \pm 0.4\%$.

Of the frictional forces acting on the $xyz$-slider, the simplest is the kinetic friction
$F_1 = \mu_k\rho g s_d$ on the moving chain segment on the table. Using the equilibrium value 
of $s_d = 6$ links, we find a result of $\mu_ks_d \approx 0.78$ link in units of $\rho g$. 
The frictional force at the table edge is $F_2 = \mu_kN$, where the normal force {\bf N} lies 
in the $xz$-plane for any angle $\psi$. (The fact that $N_y$ vanishes for an ideal table edge 
that is uniform in the $y$-direction comes about because there is no preference for either the 
positive or the negative $y$ direction.) This normal force is well approximated by a single 
force acting at a sharp table edge at an angle $\phi_N = 45^\circ$ from the upward vertical 
direction. In the absence of any friction on the table or at the table edge, $N$ is related 
to the $z$-component $T_z$ of the chain tension just above the table edge as
\begin{eqnarray}
N\cos\phi_N \approx T_z = \rho L\ddot{Z} = \rho d \frac{g}{\sqrt{8}},
\label{N}
\end{eqnarray}
where $Z$ is the $z$-coordinate of the center of mass of the chain. Hence 
$N \approx \rho dg/2$. If table and edge frictions are included in the force balance in 
the $z$-direction, the resulting expression is considerably more complicated, as we shall 
show in \ref{ChForces}. The normal force is then increased to $0.85\rho dg$, 
and yields the table edge frictional force of $F_2 \propto 0.85\mu_kd = 0.44$ link worth 
of chain weight. In the experiment, we have allowed only 0.5 link for these frictional 
forces. There is thus a shortfall of 0.72 link. Consequently the raw acceleration 
$a_{\rm raw}$ should be increased by 0.5\%.

To understand air friction $F_4$ on the sliding chain moving in air, we study its effect 
on the motion of a certain falling chain studied experimentally by Tomaszewski, 
Pieranski and Geminard \cite{TPG06} (TPG) and shown in their figure~6(a). This chain is 
a light ball chain that is initially stretched taut in the form of a horizontal flat 
catenary. One end is then released while the other end is held fixed. The falling end is 
found to be part of a horizontal chain segment that is falling freely. The length of this 
freely falling horizontal chain segment decreases until it merges completely into the 
swinging arm attached to the fixed support. The almost straight swinging arm at that 
moment has swung down from the horizontal direction by an angle of roughly 80$^\circ$. 
This is just after frame \# 22 in TPG figure 6(a) where the (vertical, horizontal) 
displacement of the chain end from its initial position is $(h, w) = (0.96, 0.76)$ m.  

To isolate the effect of air friction on this TPG falling flat chain, we note that the 
time of free fall to frame \# 22 is $t_{\rm FF} = \sqrt{2h/g} = 0.442$ s. We find from 
TPG figure 5(a) that the their theoretical falling time for that frame is smaller than 
their experimental time by $\Delta t = 0.0026 \pm 0.0003$ s. The TPG theory already 
contains bending friction between links in the chain, but not air friction. Hence the 
fractional effect of air friction is 
\begin{eqnarray}
\frac{\Delta t}{t_{\rm exp}} = 0.0059 = -\frac{\Delta a}{2a},
\label{tRatio}
\end{eqnarray}
where $a$ ($= g$ here) is the constant vertical acceleration. (We do not use the 
free-fall curve in TPG figure 5(a) to estimate air friction for the reason 
that this TPG chain actually falls slightly faster than $g$ because its initial 
position is not exactly horizontal.) 

The TPG falling flat chain is simply related to our $xy$-slider. The $xy$-slider also 
has a horizontal chain segment, the part still on the table, that diminishes in length 
as the chain falls. However, the $xy$-slider is fixed at the end of the horizontal 
segment, while the falling flat chain is fixed at the end of the swinging arm. So they 
look very much like an upside-down version of each other, but with some important 
differences. The angle $\theta$ of the swinging arm in the falling flat chain changes 
continuously, instead of being constant for a significant part of the fall. What is 
much more important to us is that these chains have different falling accelerations: 
the TPG chain falls with a vertical acceleration $g$, but the $xy$-slider falls with 
a vertical acceleration $g/3$. How does this change in acceleration affect the fractional 
change in the falling time when the retarding effect of air friction is removed from 
the experimental result?

This question is easily answered for a mass $m$ falling down vertically with constant 
acceleration $a = \ddot{x}$ that experiences an additional weak frictional force $2\beta mv$
proportional to the instantaneous velocity $v = \dot{x}$ like that acting on the 
Millikan falling oil drop. The resulting acceleration change, $\Delta a = -2\beta v$, can be
expressed as the fractional change 
\begin{eqnarray}
f(a) \equiv \frac{|\Delta a|}{a} = 2\beta t = 2\beta\sqrt{\frac{2x}{a}}.
\label{fa}
\end{eqnarray} 
Hence the fractional acceleration changes of the same mass falling through the same 
vertical distance $x$ in the same medium under two different accelerations $a_1$ and 
$a_2$ are related as
\begin{eqnarray}
\sqrt{a_2}f(a_2) = {\rm const} = \sqrt{a_1}f(a_1).
\label{faRatio}
\end{eqnarray} 
If the TPG chain falls vertically like this mass, we shall have 
\begin{eqnarray}
f(g/3) = \sqrt{3} f(g) = 0.020.
\label{fa-g3}
\end{eqnarray} 
This results holds for the $xyz$-slider also if its frictional coefficient $\beta$ is the
same as that for the TPG chain.

The TPG chain is a light ball chain, while ours is a cable chain that is shorter in length 
and smaller in mass, 0.95 m and 12 g instead of 
1.02 m and 21 g. Their cross sectional dimensions are comparable, but their cross sectional 
shapes are not the same. However, both our $xyz$-slider and the TPG flat chain are falling
at roughly the same angle to the vertical direction when air friction is strongest, i.e.,
when the falling speed is greatest towards the end of the measured fall. So their effective 
cross sections are quite similar in spite of the differences in shapes and in falling 
configurations. We conclude that their $\beta$ coefficients are likely to be quite similar, 
and that our experimental vertical acceleration should be increased by 2.0\% before 
comparison with the theoretical value for a frictionless chain.

Adding the small 0.5\% correction for the uncompensated part of the friction on the table 
and the table edge, we find a total correction of 2.5\%. The uncertainty of this correction 
is hard to estimate, but our gut feeling is that the correction is semi-realistic and that 
it is better to make it than not. We choose a relatively large uncertainty of 1.2\% in order 
to allow for the other neglected effects mentioned in this appendix. This uncertainty is 
much greater than the uncertainty of 0.4\% in the initial overhang position of the chain 
from the table edge. The latter uncertainty can thus be neglected.  

\section{Chain tension, normal and frictional forces}
\label{ChForces}

If the table is not smooth, two forces of kinetic friction act on the $xyz$-slider. They are
(a) a horizontal force acting on the moving chain segment of length $s_d = d/\sin\psi$, 
and (b) a tangential
force perpendicular to the normal force {\bf N} at the table edge. These forces will be 
referred to in the following as the table friction and the edge friction, respectively. 
We shall show in this appendix that their effect on our measured acceleration is quite 
small. Nevertheless, their proper treatment requires a certain appreciation of the dynamics 
of our falling chains that is interesting in its own right. 

\begin{figure}
\includegraphics{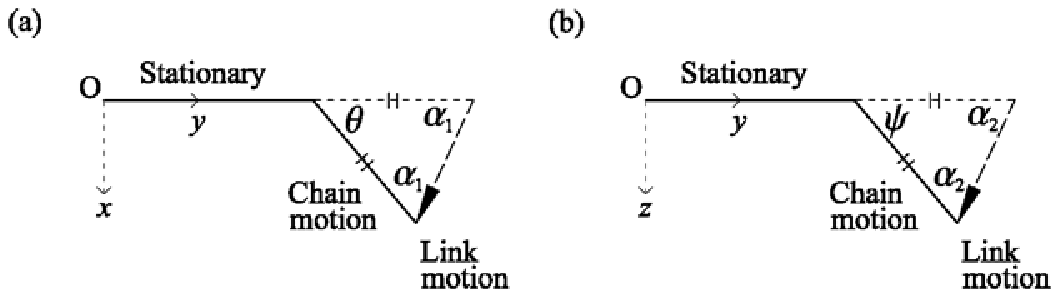}
\caption{\label{fig:Triangles}
Link and chain motions for (a) the $xy$-slider, and (b) the on-table chain segment of the 
$xyz$-slider. } 
\end{figure}
Consider first the simpler $xy$-slider whose vertically falling motion is illustrated in 
figure~\ref{fig:Triangles}(a), where the triangle shown can be of any size. The links on
the horizontal side of the triangle shown in the figure as a dotted line are originally 
part of the horizontal and stationary chain segment at the table edge. These links then 
fall along the side of the triangle shown by the directed dashed line. Hence the chain 
appears to fall along the third side of the triangle shown as a solid line 
in the figure. The triangle is thus an isosceles triangle with the apical angle $\theta$ 
defined in section~\ref{sect:xySliderMotion}, and two equal base angles
\begin{eqnarray}
\alpha_1 = {\textstyle \frac{1}{2}} (\pi - \theta).
\label{alpha_1}
\end{eqnarray}
The linear motions of links and chain occur everywhere on the falling chain if the 
apical angle $\theta$ remains constant in time. 

Very much the same situation also appears on the table for the setback $xyz$-slider. Here a 
short chain segment of length $s_d$ is sliding on the table with the links moving along 
the directed dashed side of the triangle shown in figure~\ref{fig:Triangles}(b), while the 
chain itself appears to move along the solid side of the triangle. The triangle has an 
apical angle $\psi$ and two equal base angles
\begin{eqnarray}
\alpha_2 = {\textstyle \frac{1}{2}} (\pi - \psi).
\label{alpha_2}
\end{eqnarray}
The table friction in the laboratory frame is thus opposite in direction to the directed 
dashed side of the triangle .

We now turn to the friction at the table edge. Experience shows that the chain direction 
defined by $\psi \approx 45^\circ$ actually persists for some distance around and below 
the table edge before the overhang angle $\theta \approx 70^\circ$ finally takes over. 
Thus one may visualize figure~\ref{fig:Triangles}(b) to be folded around the table edge 
and to continue for some distance below it before the apical angle changes to $\theta$. 
The plane containing the edge friction at the sharp table edge can be obtained by 
first placing figure~\ref{fig:Triangles}(b) on the table top and then rotating it about 
the table edge by an angle $\phi_N \approx 45^\circ$ until its normal direction coincides 
with the direction of the normal force {\bf N}. Edge friction is then opposite in direction 
to the directed dashed side of that rotated isosceles triangle whose lower base angle is 
located at the point where {\bf N} acts. Just for the sake of notational transparency, 
we shall denote this new base angle $\alpha_3$.

We are now in a position to decompose forces along different laboratory coordinate axes. 
For the $z$-components at the table edge of the $xyz$-slider, we find
\begin{eqnarray}
N\cos\phi_N - \mu_kN\cos\phi_N\sin\alpha_3 &-& \mu_k\rho s_d g \sin\alpha_2 
\nonumber \\
&=& T_z = \rho L\ddot{Z} = \rho d\ddot{s}
\label{zForces}
\end{eqnarray}
for a chain whose center-of-mass position is $(X, Y, Z)$. This equation
can be solved for the normal force:
\begin{eqnarray}
N &=& \frac{\rho dg}{\cos\phi_N(1-\mu_k\sin\alpha_3)} 
\left(\frac{\ddot{s}}{g} + \mu_k\frac{\sin\alpha_2}{\sin\psi} \right) \\
&\approx& 0.85 \rho dg,
\label{NForces}
\end{eqnarray}
where the numerical value is calculated for $\mu_k =0.13$, $\psi = 45^\circ$, and 
$\alpha_3 = \alpha_2$.

As another example, the chain tension acting on the fixed chain end at the origin O of the
$xyz$-slider is
\begin{eqnarray}
T &=& T_y = \rho L\ddot{Y} + \mu_kN\cos\alpha_3 + \mu_k\rho d_sg\cos\alpha_2, 
\label{TensionO}
\end{eqnarray}
where the terms can be calculated in terms of $\ddot{s}$ and the various parameters. 
The numerical result is $T \approx 0.71 \rho sg$.


\section*{References}
\begin{thebibliography}{32}

\bibitem{Wong06} Wong C W and Yasui K 2006 
Falling chains {\it Am. J. Phys.}  {\bf 74} 490--496 
arXiv: physics/0508005

\bibitem{Cayley57} 
Cayley A 1857 On a class of dynamical problems {\it Proc. Roy. Soc. London} 
{\bf 8} 506--511 Reprinted in Ref.~\cite{Cayley89} Vol~IV pp~7--11 

\bibitem{Irschik04} 
Irschik H and Holl H J 2004 Mechanics of variable-mass systems---Part 1: 
Balance of mass and linear momentum {\it App. Mech. Rev.} {\bf 57} 
145--160  

\bibitem{Villarino06} 
Villarino M B 2006 email communications 30 January

\bibitem{Routh98} 
Routh E J 1898 \textsl{A Treatise on Dyamics of a Particle} (Cambridge: Cambridge 
University Press) pp~80--81

\bibitem{Tait56} 
Tait P G and Steele W J 1856 \textsl{A Treatise on Dyamics of a Particle} 
(london: MacMillan \& Co) pp~250--251; 1900 7th edn pp~334--335

\bibitem{Mikhailov84} 
Mikhailov G K 1984 The dynamics of mechanical systems with variable masses 
as developed at Cambridge during the second half of the nineteenth century 
{\it Bull. Inst. Math. Appl.} {\bf 20} 13--19

\bibitem{Warwick03} 
Warwick A 2003 \textsl{Masters of Theory, Cambridge and the Rise of Mathematical 
Physics} (Chicago: University of Chicago Press) pp~157--158

\bibitem{camcorder}
Sony Model DCR-HC65 NTSC Serial No. 1327219

\bibitem{Sommerfeld52} 
Sommerfeld A 1952 \textsl{Lectures on Theoretical Physics, Vol.~I, Mechanics} 
(New York: Academic Press) pp~28--29, Problem~I.7 p~241, and solution p~257

\bibitem{Ball60} 
Rouse Ball W W 1960 \textsl{A Short Account of the History of Mathematics} 
(New York: Dover) p~428 

\bibitem{Gillispie71} 
Gillispie C C 1971 \textsl{Lazare Carnot Savant} 
(Princeton: Princeton University Press) p~57, where an English
translation is given of the relevant sentences from Carnot L 1783
\textsl{Essai sur les machines} (Dijon, Burgundy) Article~LIX pp~91--92 

\bibitem{Cayley57BA}
Cayley A 1857 Report on the recent progress of theoretical dynamics 
{\it Brit. Assoc. Report} 1--42 Reprinted in 
Ref.~\cite{Cayley89} Vol~III pp~156--204 

\bibitem{Crilly06} 
Crilly A J 2006, \textsl{Arthur Cayley: Mathematician Laureate of the Victorian age} 
(Baltimore: Johns Hopkins) pp~222--3  

\bibitem{Warwick03a} 
Warwick A \cite{Warwick03} p~89

\bibitem{Warwick03b} 
Warwick A \cite{Warwick03} pp~87 231

\bibitem{Forsyth95}
Forsyth A R 1895 Arthur Cayley {\it Obit. Notices, Proc. Roy. Soc. London} 
i--xliii 

\bibitem{Crilly98}
Crilly T 1998 The young Arthur Cayley {\it Notes and Records Roy. 
Soc. London} {\bf 52} 267--282

\bibitem{Wolstenholme78} 
Wolstenholme J \textsl{Mathematical Problems}, 2nd edn (London: Macmillan \& Co) 
(We have access only to the 3rd edition of 1891 where problems related 
to the HTSC falling chain problem appear as Problems 2354 and 2355 on p~419.
A notice in the book states that the third edition is essentially a reprint of the 
second edition with most of the errors corrected.)

\bibitem{Jeans07} 
Jeans J M 1907 \textsl{An Elementary Treatise on the Theoretical Mechanics} 
(Boston: Ginn and Co) pp~236--237  

\bibitem{Lamb14} 
Lamb H 1914 \textsl{Dynamics} 
(Cambridge: Cambridge University Press) pp~143--144  

\bibitem{Warwick03e} 
Warwick A \cite{Warwick03} pp~156 260 514 520 

\bibitem{Sony} 
Sony Electronics has declined to provide any information on their timer
accuracy because their `engineering data is proprietary'.

\bibitem{Wolfram} 
Wolfram S 1999 \textsl{The Mathematica Book} 4th edn (Wolfram Media/Cambridge University 
Press)

\bibitem{Rosenfeld75} Rosenfeld A H 1975 The Particle Data Group: Growth and 
Operations-Eighteen Years of Particle Physics
{\it Ann. Rev. Nucl. Sci.}  {\bf 25} 555--598 (The use of scale factors is described on 
p~580.) 

\bibitem{HCP91} Lide D R (ed) 1991 {\it Handbook of Chemistry and Physics} 72 edn
(Boca Raton: CRC Press) p~14-7

\bibitem{Wong06a} Wong C W, Yasui K and Youn S H unpublished

\bibitem{TeflonTape}
Uline 3 mil Teflon tape Catalog No. S-7196 uline.com

\bibitem{Persson00} 
Persson B N J 2000 \textsl{Sliding friction} (Berlin: Springer) pp~9--19  

\bibitem{RoyMech} 
http://www.roymech.co.uk/Useful\_Tables/Tribology/co\_of\_frict.htm

\bibitem{TPG06} Tomaszewski W, Pieranski P and Geminard J-C 2006 
The motion of the freely falling chain tip {\it Am. J. Phys.} {\bf 74} 776--783  
arXiv: physics/0510060

\bibitem{Cayley89}
Cayley A 1989--97 \textsl{The Collected Mathematical Papers of Arthur Cayley} 
(Cambridge: Cambridge University Press)   


\end{thebibliography}
\end{document}